\documentclass[review,12pt]{elsarticle}

\usepackage{amssymb}
\usepackage{amsthm}
\usepackage{amsmath}
\usepackage{mathrsfs}
\usepackage{graphicx}
\usepackage{epstopdf}
\usepackage{float}
\usepackage{caption}
\usepackage{subcaption}
\usepackage{bm}
\usepackage{bbm}
\usepackage{mathrsfs}
\usepackage{cleveref}
\usepackage{soul}
\usepackage{multirow}
\usepackage{xcolor}
\usepackage{framed} 
\usepackage{nomencl} 
\makenomenclature 
\biboptions{sort&compress}

\journal{Acta Materialia}

\makeatletter
\def\@author#1{\g@addto@macro\elsauthors{\normalsize%
    \def\baselinestretch{1}%
    \upshape\authorsep#1\unskip\textsuperscript{%
      \ifx\@fnmark\@empty\else\unskip\sep\@fnmark\let\sep=,\fi
      \ifx\@corref\@empty\else\unskip\sep\@corref\let\sep=,\fi
      }%
    \def\authorsep{\unskip,\space}%
    \global\let\@fnmark\@empty
    \global\let\@corref\@empty  
    \global\let\sep\@empty}%
    \@eadauthor={#1}
}
\makeatother

\begin{document}

\begin{frontmatter}



\title{Analysis of the influence of microstructural traps on hydrogen assisted fatigue}


\author{Rebeca Fern\'{a}ndez-Sousa \fnref{Uniovi}}
\author{Covadonga Beteg\'{o}n \fnref{Uniovi}}
\author{Emilio Mart\'{\i}nez-Pa\~neda\corref{cor1}\fnref{IC}}
\ead{e.martinez-paneda@imperial.ac.uk}

\address[Uniovi]{Department of Construction and Manufacturing Engineering, University of Oviedo, Gij\'{o}n 33203, Spain}

\address[IC]{Department of Civil and Environmental Engineering, Imperial College London, London SW7 2AZ, UK}

\cortext[cor1]{Corresponding author.}

\begin{abstract}
We investigate the influence of microstructural traps on hydrogen diffusion and embrittlement in the presence of cyclic loads. A mechanistic, multi-trap model for hydrogen transport is developed, implemented into a finite element framework, and used to capture the variation of crack tip lattice and trapped hydrogen concentrations as a function of the loading frequency, the trap binding energies and the trap densities. We show that the maximum value attained by the lattice hydrogen concentration during the cyclic analysis exhibits a notable sensitivity to the ratio between the loading frequency and the effective diffusion coefficient. This is observed for both hydrogen pre-charged samples (closed-systems) and samples exposed to a permanent source of hydrogen (open-systems). Experiments are used to determine the critical concentration for embrittlement, by mapping the range of frequencies where the output is the same as testing in inert environments. We then quantitatively investigate and discuss the implications of developing materials with higher trap densities in mitigating embrittlement in the presence of cyclic loads. It is shown that, unlike the static case, increasing the density of “beneficial traps” is a viable strategy in designing alloys resistant to hydrogen assisted fatigue for both closed- and open-systems.
\end{abstract}


\end{frontmatter}


\begin{framed}
\nomenclature{$a$}{crack length}
\nomenclature{$b$}{Burgers vector}
\nomenclature{$C$}{total hydrogen concentration}
\nomenclature{$C_L, \, C_T$}{hydrogen concentration in lattice and trapping sites}
\nomenclature{$C_0$}{initial hydrogen concentration}
\nomenclature{${C_L}_{max,\theta=0^\circ}, \, {C_T}_{max,\theta=0^\circ} , \, {C}_{max,\theta=0^\circ}$}{maximum value of the lattice, trapped and total hydrogen concentrations attained at any material point along the extended crack plane ($r, \theta=0^\circ$) for a specific time instant}
\nomenclature{${C_L}_{max,N}$}{maximum value of the lattice hydrogen concentration attained within each cycle along the extended crack plane ($r, \theta=0^\circ$)}
\nomenclature{$D, D_e$}{lattice and effective diffusion coefficients}
\nomenclature{$D_g$}{mean grain size}
\nomenclature{$d_j$}{diameter of carbide particle $j$}
\nomenclature{$E$}{Young's modulus}
\nomenclature{$f$}{load frequency}
\nomenclature{$\bm{J}$}{hydrogen flux}
\nomenclature{$K_T^{(i)}$}{equilibrium constant for the $i$th type of trapping sites}
\nomenclature{$\Delta K$}{stress intensity factor amplitude}
\nomenclature{$K_{min}, \, K_m, \, K_{max}$}{minimum, mean and maximum stress intensity factor}
\nomenclature{$\ell$}{material gradient length scale}
\nomenclature{$L$}{average distance between the carbide particles}
\nomenclature{$M$}{Taylor's factor}
\nomenclature{$N$}{number of cycles}
\nomenclature{$\mathcal{N}$}{strain hardening exponent}
\nomenclature{$N_A$}{Avogadro's number}
\nomenclature{$N_L$}{number of lattice sites per unit volume}
\nomenclature{$N_T^{(c)}$}{number of carbide trapping sites per unit volume}
\nomenclature{$N_T^{(d)}$}{number of dislocation trapping sites per unit volume}
\nomenclature{$N_T^{(m)}$}{number of martensitic interfaces trapping sites per unit volume}
\nomenclature{$\mathcal{R}$}{universal gas constant}
\nomenclature{$R$}{load ratio}
\nomenclature{$R_p$}{reference size of the plastic zone}
\nomenclature{$r_0$}{initial crack tip blunting radius}
\nomenclature{$\bar{r}$}{Nye's factor}
\nomenclature{$r$,\, $\theta$}{polar coordinates}
\nomenclature{$T$}{absolute temperature}
\nomenclature{$u , \, v$}{horizontal and vertical components of the displacement field}
\nomenclature{$\bar{V}_H$}{partial molar volume of hydrogen}
\nomenclature{$V_M$}{molar volume of the host lattice}
\nomenclature{$W_B^{(i)}$}{binding energy for the $i$th type of trapping sites}
\nomenclature{$\beta$}{number of lattice sites per solvent atom}
\nomenclature{$\varepsilon^p$}{equivalent plastic strain}
\nomenclature{$\eta^p$}{effective plastic strain gradient}
\nomenclature{$\theta_L, \, \theta_T^{(i)}$}{occupancy of lattice and $i$th type of trapping sites}
\nomenclature{$\mu$}{shear modulus}
\nomenclature{$\nu$}{Poisson's ratio}
\nomenclature{$\rho$}{dislocation density}
\nomenclature{$\sigma_H$}{hydrostatic stress}
\nomenclature{$\sigma_f$}{tensile flow stress}
\nomenclature{$\tau$}{shear flow stress}
\printnomenclature
\end{framed}

\section{Introduction}
\label{Sec:Intro}

Hydrogen originating from water vapour, aqueous electrolytes or gaseous environments significantly increases cracking susceptibility and fatigue crack growth rates in metals \cite{Gangloff2003,Gangloff2012}. As a consequence, there is an increasing interest in developing reliable prognosis methodologies based on a mechanistic understanding of this so-called hydrogen embrittlement phenomenon. In this realm, efforts include insightful experimentation \cite{Wang2014,Girardin2015,Harris2018,Nagumo2019} and the development of theoretical and numerical models for hydrogen transport \cite{Sofronis1989,Krom1999,CS2020b}, fracture \cite{Kirchheim2015,Nagao2018,CMAME2018,Tehranchi2019,Shishvan2020,JMPS2020} and fatigue \cite{Moriconi2014,EFM2017}; see Ref. \cite{Djukic2019} for a comprehensive review.\\

Hydrogen atoms can reside at interstitial lattice sites and microstructural trapping sites, such as dislocations, grain boundaries, voids, carbides and interfaces \cite{Hirth1980,Pressouyre1979}. Traps act as hydrogen sinks, slowing diffusion, and are typically characterised by their binding energy $W_B$ and density $N_T$. The energy barrier that must be overcome for the hydrogen to detrap increases with $|W_B|$; hydrogen will be strongly retained in deep traps $|W_B|>60$ kJ/mol but can be easily released from shallow traps $|W_B|<30$ kJ/mol. Quantifying this partitioning of hydrogen atoms between lattice and trapping sites is of utmost importance in predicting diffusion and embrittlement; see, e.g.,  \cite{Li2004,Pundt2006a,Novak2010,Turnbull2015} and references therein. Moreover, understanding the interaction of multiple trap states with diffusible hydrogen is a key step in imbuing materials with intrinsic resilience \cite{Spencer1998,Yamasaki2006,Bhadeshia2016,Chen2017,Breen2020,Chen2020}. The ambition is to design hydrogen embrittlement-resistant alloys by engineering microstructures with a high density of \emph{beneficial} traps, which will retain the hydrogen and hinder diffusion to the fracture process zone. One strategy involves incorporating finely dispersed nano-scale carbides \cite{Ramjaun2018,Turk2018}. Vanadium carbides have been successfully used to mitigate hydrogen embrittlement in refinery pressure vessels and other \emph{closed-systems}, where hydrogen entry is essentially a one-off process. However, the works by Dadfarnia \textit{et al.} \cite{Dadfarnia2011} and Hosseini \textit{et al.} \cite{Hosseini2017} have shown that increasing the density of traps to sequester hydrogen is not a viable strategy to mitigate embrittlement in \emph{open-systems}, where there is a permanent source of hydrogen. Increasing the density of one type of trap decreases the effective diffusion of hydrogen and delays the time required to achieve the steady state but has no effect on the content of hydrogen in the lattice or any other type of trap once the steady state is reached. Moreover, even for high trap densities, the steady state is attained in days, with the lattice hydrogen reaching 98\% of its steady state magnitude in minutes - a very short time frame relative to the lifetime of an engineering component \cite{Dadfarnia2011}. But notably, the analyses of Dadfarnia \textit{et al.} \cite{Dadfarnia2011} and Hosseini \textit{et al.} \cite{Hosseini2017} are limited to monotonic/static loading conditions. In fatigue, each loading cycle is significantly faster than the time required to achieve steady state and experiments show that embrittlement is precluded if the ratio between the loading frequency and the (effective) diffusion coefficient is sufficiently low \cite{Murakami2010a,Fassina2013,Tazoe2017,Alvaro2019,Peral2019}.\\

In this work, we combine numerical analysis and experimental data to gain insight into the influence of microstructural traps in hydrogen assisted fatigue. A multi-trap model based on Oriani's equilibrium \cite{Oriani1974} and Taylor's dislocation model \cite{Taylor1938} is developed and used to analyse conditions relevant to both open and closed-systems. We shed light into the competition between multiple types of traps in governing hydrogen diffusion ahead of fatigue cracks and reveal that increasing the density of a specific trap type is a viable strategy for extending the range of loading frequencies where embrittlement is not observed. In addition, the influence of crack tip plastic strain gradients is incorporated into the modelling of multi-trap systems and hydrogen assisted fatigue for the first time. 

\section{Theory}
\label{Sec:Theory}

\subsection{Multi-trap model for hydrogen transport}
\label{Sec:hydrogenTransportModel}

Denote the lattice hydrogen concentration as $C_L$, which is given by
\begin{equation}\label{eq:CL}
    C_L=\theta_{L} N_L,
\end{equation}

\noindent where $N_L$ is the number of lattice sites per unit volume and $\theta_L$ is the lattice site occupancy. The former is a function of the molar volume of the host lattice $V_M$, the number of interstitial sites per solvent atom $\beta$ and Avogadro's number $N_A$ as $N_L=\beta N_A/V_M$. The choice of $\beta=6$ (as for bcc, see Ref. \cite{Krom1999}) leads to $N_L=5.1 \times 10^{29}$ sites/m$^3$ \cite{Sofronis1989,DiLeo2013}. On the other hand, the trapped hydrogen concentration for the $i$th type of trapping sites is given by
\begin{equation}\label{eq:CT}
    C_T^{(i)} = \theta_T^{(i)} N_T^{(i)},
\end{equation}

\noindent where $N_T$ is the trap density (trapping sites per unit volume) and $\theta_T$ is the fraction of occupied trapping sites.\footnote{Alternatively, one can define $C_T=\theta_T \alpha N_T$, with $N_T$ being the number of traps per unit volume and $\alpha$ the number of atom sites per trap. Since commonly $\alpha=1$ \cite{Krom1999}, we choose to denote the trap density as the number of trapping sites per unit volume $N_T \equiv \alpha N_T$.} The trap density is a material property that remains constant throughout the analysis for traps such as carbides or grain boundaries but that evolves with mechanical loading for the case of dislocation traps; a Taylor-based formulation is presented below to determine $N_T^{(d)}$. We adopt Oriani's equilibrium theory \cite{Oriani1974}, resulting in the following Fermi-Dirac relation between the occupancy of the $i$th type of trapping sites and the fraction of occupied lattice sites
\begin{equation}\label{eq:Oriani}
    \frac{\theta_T^{(i)}}{1 - \theta_T^{(i)}} = \frac{\theta_L}{1- \theta_L} K_T^{(i)},
\end{equation}

\noindent with $K_T^{(i)}$ being the equilibrium constant for the $i$th type of trap with binding energy $W_B^{(i)}$; given by
\begin{equation}\label{eq:KT}
    K_T^{(i)}=\exp \left( \frac{-W_B^{(i)}}{\mathcal{R}T} \right).
\end{equation}

\begin{figure}[H]
  \makebox[\textwidth][c]{\includegraphics[width=0.9\textwidth]{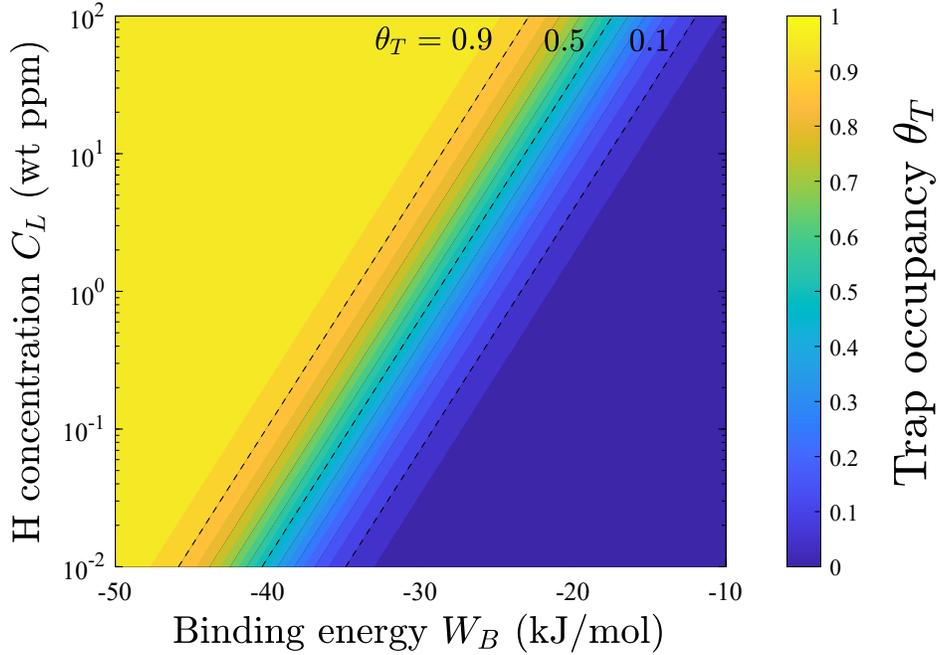}}
  \caption{Implications of Oriani's equilibrium; sensitivity of the trap occupancy $\theta_T$ to the lattice hydrogen concentration $C_L$ and the trap binding energy $W_B$.}
  \label{fig:Oriani}
\end{figure}

\noindent Here, $\mathcal{R}=8.3145$ J/(mol$\cdot$K) is the universal gas constant and $T$ is the absolute temperature. The implications of Oriani's equilibrium are illustrated in Fig. \ref{fig:Oriani} by combining (\ref{eq:CL}), (\ref{eq:Oriani}) and (\ref{eq:KT}) to plot the contours of trap occupancy $\theta_T$ as a function of the trap binding energy $W_B$ and the lattice hydrogen concentration $C_L$. It is observed that traps with binding energies $W_B<-50$ kJ/mol saturate at very low $C_L$ values, increasing the $C_T/C_L$ ratio for a given trap density. On the other hand, shallow traps with binding energies larger than $-20$ kJ/mol are effectively empty ($\theta_T \approx 0$) unless $C_L$ is very high, on the order of 10 wt ppm ($4.68 \times 10^{25}$ at H/m$^3$) or higher.

Finally, upon the common assumption of low occupancy conditions $\theta_L << 1$, an effective diffusion coefficient can be defined by,
\begin{equation}\label{eq:De}
    D_e = D \frac{C_L}{C_L+ \sum_i C_T^{(i)} \left( 1 - \theta_T^{(i)} \right)} \, ,
\end{equation}

\noindent and the hydrogen transport equation reads
\begin{equation}
        \frac{D}{D_e}\frac{\partial C_L}{\partial t}=D \nabla^2 C_L-\nabla \left( \frac{D C_L}{\mathcal{R}T} \bar{V}_H \nabla \sigma_H  \right ) ,
    \label{Dif}
\end{equation}

\noindent where $\bar{V}_H$ is the partial molar volume of hydrogen in solid solution and $\sigma_H$ is the hydrostatic stress.

\subsection{A Taylor-based model for plastic flow and trapping}
\label{Sec:MSG}

Capturing how the trap density for dislocations $N_T^{(d)}$ evolves with the applied load requires estimating the dislocation density $\rho$. Predicting the dislocation density \textit{via} a micromechanics approach is also relevant for providing a more precise description of crack tip deformation \cite{Hutchinson1997,Komaragiri2008,IJP2016}. Here, we follow Taylor's \cite{Taylor1938} dislocation model and accordingly relate the shear flow stress to $\rho$, the shear modulus $\mu$ and the Burgers vector $b$ as
\begin{equation}\label{eq:tau}
\tau = 0.5 \mu b \sqrt{\rho}.
\end{equation}

\noindent The dislocation density $\rho$ comprises the sum of the density $\rho_{SSD}$ for statistically stored dislocations (SSDs) and the density $\rho_{GND}$ for geometrically necessary dislocations (GNDs):
\begin{equation}\label{eq:rho}
   \rho=\rho_{SSD}+\rho_{GND} .
\end{equation}

\noindent The GND density is defined by:
\begin{equation}\label{eq:rhoG}
    \rho_{GND} = \bar{r} \frac{\eta^p}{b},
\end{equation}

\noindent where $\bar{r}$ is the Nye-factor and $\eta^p$ is the effective plastic strain gradient, which is defined as follows \cite{Gao1999,IJSS2015}:
\begin{equation}
    \eta^p = \sqrt{\frac{1}{4} \eta_{ijk}^p \eta_{ijk}^p} \,\,\,\, \text{with} \, \, \eta_{ijk}^p=\varepsilon_{ik,j}^p + \varepsilon_{jk,i}^p - \varepsilon_{ij,k}^p \, ,
\end{equation}
\noindent where $\varepsilon_{ij}^p$ is the plastic strain tensor. 
The tensile flow stress $\sigma_{f}$ is proportionally related to $\tau$ \textit{via} the Taylor factor $M$ such that, considering (\ref{eq:tau})-(\ref{eq:rhoG}), 
\begin{equation}\label{eq:sigma_f}
    \sigma_{f} = M \tau =0.5 M \mu b \sqrt{\rho_{SSD} + \bar{r} \frac{\eta^p}{b}}.
\end{equation}

\noindent Here, $M=2.9$ for bcc metals. The SSD density $\rho_{SSD}$ can be determined from (\ref{eq:sigma_f}) knowing the relation in uniaxial tension $(\eta = 0)$ between the flow stress and the material stress-strain curve as follows
\begin{equation}\label{eq:rhoS}
    \rho_{SSD} = \left( \frac{\sigma_{ref} f \left( \varepsilon^p \right)}{0.5 M \mu b} \right)^2,
\end{equation}

\noindent where $\sigma_{ref}$ is a reference stress and $f(\varepsilon^p)$ is a non-dimensional function determined from the uniaxial stress-strain curve. Substituting back into (\ref{eq:sigma_f}), one reaches
\begin{equation}\label{eq:sF_msg}
 \sigma_f = \sigma_{ref} \sqrt{f^2 \left( \varepsilon^p \right) + \ell \eta^p}
\end{equation}

\noindent where $\ell$ is the intrinsic material length. If the length parameter is set to zero or $f^2 \left( \varepsilon^p \right)$ outweighs the GND contribution $\ell \eta^p$, the model recovers conventional von Mises plasticity. For the sake of clarity, we have chosen to show results first for the case of conventional plasticity ($\rho=\rho_{SSD}$, $\ell=0$) and assess later the implications of accounting for the role of plastic strain gradients. This allows for validating the coupled deformation-diffusion model for multi-trapping with the static results of Dadfarnia \textit{et al.} \cite{Dadfarnia2011} (not shown). In both conventional and strain gradient plasticity models, relating the dislocation density with macroscopic quantities such as $\varepsilon^p$ and $\eta^p$ will allow us to estimate the evolution of the dislocation trap density $N_T^{(d)}$.

\section{Methodology}
\label{Sec:Met}

\subsection{Experiments}
\label{Sec:Material}

We build our analysis in the experimental characterisation of the fatigue behaviour of hydrogen pre-charged 42CrMo4 steel samples \cite{Peral2019}. The 42CrMo4 steel under consideration was austenitized at 845$^\circ$C for 40 min, quenched in water and tempered at 700$^\circ$C for two hours. As described elsewhere \cite{Zafra2018}, the mechanical properties of the material are obtained from uniaxial tension tests, giving a yield stress of $\sigma_y=622$ MPa. The material work hardening is characterised by means of an isotropic hardening power law:
\begin{equation}
    \sigma = \sigma_y \left( 1 + \frac{E \varepsilon^p}{\sigma_y} \right)^\mathcal{N}
\end{equation}

\noindent where $\mathcal{N}=0.1$ is the hardening coefficient and $\varepsilon^p$ is the effective plastic strain. The reference stress in Eq. (\ref{eq:sF_msg}) will correspond to $\sigma_{ref}=\sigma_y (E / \sigma_y)^\mathcal{N}$ and $f \left( \varepsilon^p \right)= \left( \varepsilon^p + \sigma_y / E \right)^\mathcal{N}$.
The Young's modulus equals $E=220$ GPa and Poisson's ratio is $\nu=0.3$. Permeation tests are used to determine trap binding energies \cite{Zafra2020}, resulting in three values that are assigned to dislocations, carbides and martensitic interfaces - see Table \ref{tab:energy}. We proceed then to estimate the trap densities for each trap type. First, following Taha and Sofronis \cite{Taha2001} in assuming one trap site per atomic plane threaded by a dislocation, the dislocation trap density is given by
\begin{equation}\label{eq:N_T}
    N_T^{(d)}=\frac{\sqrt{2} \rho}{a}
\end{equation}

\noindent where $a=0.2867$ nm is the lattice parameter \cite{Nagao2018}. Since $\rho$ is defined as a function of plastic strains and plastic strain gradients, see Section \ref{Sec:MSG}, an initial dislocation trap density $N_{T,0}^{(d)}$ is defined for the unstressed state. Assuming a recrystallised microstructure, $N_{T,0}^{(d)}$ can be estimated \textit{via} (\ref{eq:N_T}) from an initial dislocation density of $\rho_0=10^{14}$ m$^{-2}$. In regards to the trap density for carbides, we follow Nagao \textit{et al.} \cite{Nagao2018} and infer the volume density of carbide particles and the number of hydrogen trap sites per particle from SEM micrographs. Namely, the volume density is given by $(1/L^3)$ with $L=125$ nm being the average distance between the carbide particles, and the trap site density for carbide sites can then be estimated from each carbide particle diameter $d_j$ and associated frequency $f_j$ as follows:
\begin{equation}\label{eq:NTcarb}
    N_T^{(c)}=\left(\sum_j \pi d_j^2 f_j \right) \frac{4}{a^2} \frac{1}{L^3}
\end{equation}

\noindent The value obtained is listed in Table \ref{tab:energy}; the sensitivity to $N_T^{(c)}$ will be explored, as increasing the carbide content is the main strategy in designing materials with \emph{beneficial} traps and intrinsic resilience \cite{Ramjaun2018,Turk2018}.\\

Finally, the trap density associated with martensitic interfaces, $N_T^{(m)}$, is estimated following the work by Galindo-Nava \textit{et al.} \cite{Galindo-Nava2017}. Thus, $N_T^{(m)}$ can be given as a function of the Burgers vector, the lattice site density and the mean grain size $D_g=2.5$ $\mu$m as
\begin{equation}
    N_{T}^{(m)}=\frac{b}{D_g}N_L
    \label{NTint}
\end{equation}

Martensitic interfaces are assumed to have a similar size to that of the mean grain size, a simplification that would allow for an alternative interpretation of the permeation data. Thus, if lath boundaries were to be of low misorientation and difficult to distinguish from dislocations in the context of permeation and desorption data, one could effectively re-interpret $N_{T}^{(m)}$ as the trap density of prior austenite grain boundaries. The binding energies and trap densities for each trap type are listed in Table \ref{tab:energy}, where the trap density for dislocations corresponds to that of the unstressed state, $N_{T,0}^{(d)}$. 

\begin{table}[H]
  \centering
  \caption{Binding energies $W_B$ and trap densities $N_T$ measured for 42CrMo4 steel. The trap density for dislocations corresponds to that of the unstressed state, $N_{T,0}^{(d)}$}
    \begin{tabular}{|c|c|c|}
    \hline
    \textbf{Trap type} & \textbf{$W_B$ [kJ/mol]} & \multicolumn{1}{c|}{\textbf{$N_T$ [sites/m$^{3}$]}} \\
    \hline
    Dislocations & -35.2  & $4.93 \times 10^{23}$ \\
    Carbides & -21.4  & $3.61 \times 10^{23}$ \\
    Martensitic interfaces & -24.7 & $5.06 \times 10^{25}$ \\
    \hline
    \end{tabular}
  \label{tab:energy}
\end{table}

\begin{figure}[H]
  \makebox[\textwidth][c]{\includegraphics[width=0.9\textwidth]{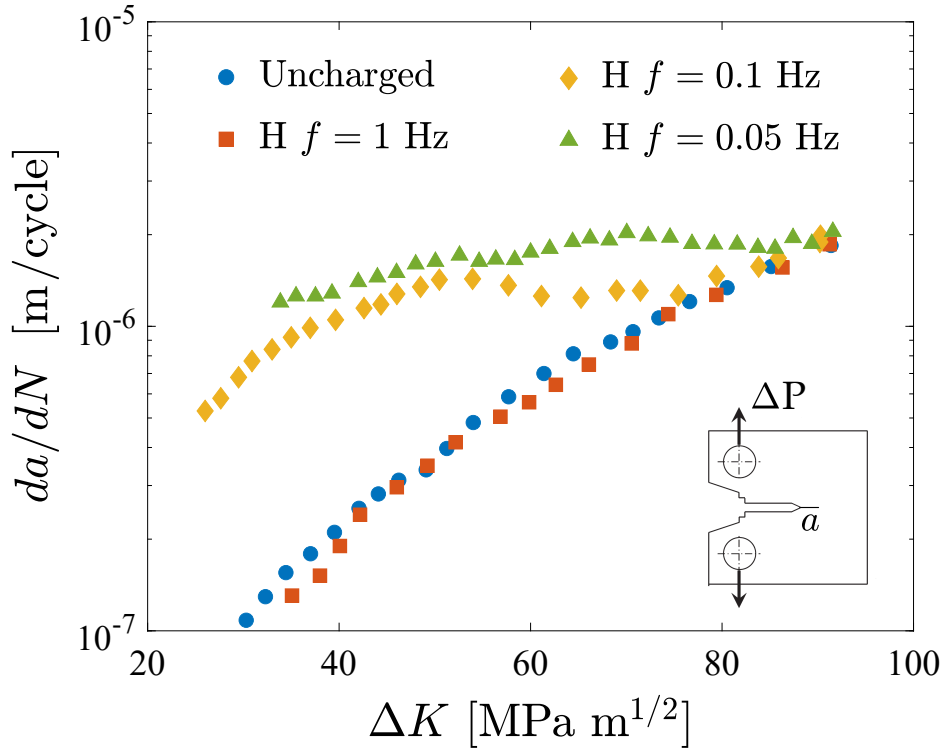}}
  \caption{Experimental results in 42CrMo4 steel subjected to a load ratio of $R=0.1$ \cite{Peral2019}. Crack growth rates $da/dN$ versus load amplitude $\Delta K$. It is shown that there is a frequency threshold below which hydrogen has no effect on the fatigue behaviour.}
  \label{fig:Propagation}
\end{figure}

The lattice diffusion coefficient is measured using permeation and estimated to be $D_L=1.3 \times 10^{-9}$ m$^2$/s \cite{Zafra2020}. All the specimens are pre-charged with gaseous hydrogen in a high-pressure hydrogen reactor for 21 h at 450$^\circ$C under a pressure of 19.5 MPa of pure hydrogen to ensure that the samples are saturated with hydrogen (10 mm thickness) \cite{Peral2019}. Thermal Desorption Spectroscopy (TDS) is used in combination with diffusion modelling to estimate the initial lattice hydrogen concentration, which equals $C_0=1.06$ wt ppm ($4.96 \times 10^{24}$ at H/m$^3$). The fatigue crack growth experiments were conducted using compact tension (CT) specimens with a width of 48 mm and a thickness of 10 mm, see Ref. \cite{Peral2019} for details. Before hydrogen pre-charging, the samples were first fatigue pre-cracked at a load ratio of $R=0.1$ and 10 Hz until reaching a crack length to width ratio of $a/W=0.2$, following the ASTM E647 standard. The results obtained in both uncharged and pre-charged samples loaded at different frequencies are shown in Fig. \ref{fig:Propagation} in terms of crack growth rates $da/dN$ versus load amplitude $\Delta K$. A load ratio of $R=\Delta K_{min} / \Delta K_{max}=0.1$ is used and experiments are conducted at room temperature. The experimental results reveal that the behaviour of the hydrogen-free samples is recovered in the hydrogen-charged experiments if the loading frequency is higher than 1 Hz. For lower frequencies, hydrogen embrittles the material and accelerates crack growth rates. The existence of a \emph{safe} regime of loading frequencies, where hydrogen has no effect, has also been demonstrated in other experimental works \cite{Murakami2010a,Fassina2013,Tazoe2017,Alvaro2019}.

\subsection{Numerical model}
\label{Subsec:FE}

\begin{figure}[H]
  \makebox[\textwidth][c]{\includegraphics[width=1.2\textwidth]{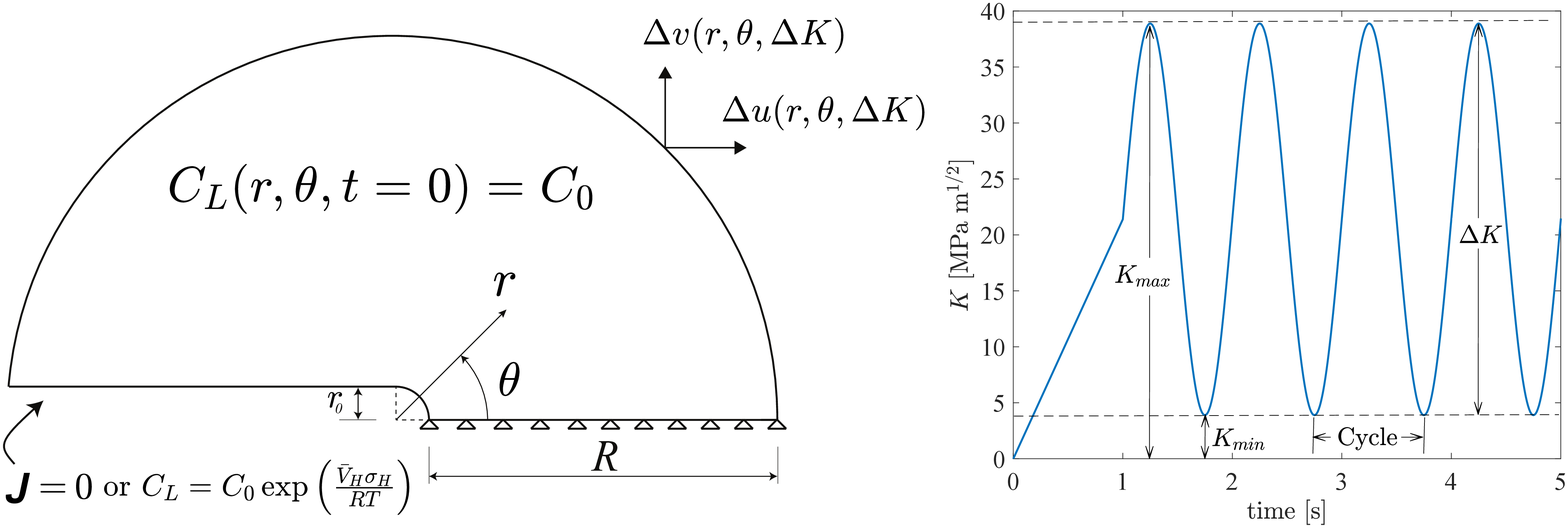}}
  \caption{Sketch of the numerical model: boundary layer formulation, with mechanical and diffusion boundary conditions, and applied $K$ as a function of time for the case of $R=0.1$, $\Delta K=35$ MPa$\sqrt{m}$, and $f=1$ Hz.}
  \label{fig:Boundary}
\end{figure}

Hydrogen transport during cyclic loading and its implications for embrittlement are investigated using a finite element model. Both qualitative and quantitative insight is gained, using the fatigue experiments on 42CrMo4 steel by Peral \textit{et al.} \cite{Peral2019} for the latter (see Section \ref{Sec:Material}). Small scale yielding conditions apply and accordingly crack tip fields are computed by using a boundary layer formulation \cite{Sofronis1989}. Hence, as described in Fig. \ref{fig:Boundary}, the crack region is contained by a circular zone and a \emph{cyclic} remote Mode I load $\Delta K$ is applied by prescribing the horizontal $u$ and vertical $v$ displacement components of the nodes at the remote circular boundary:
\begin{equation}
   \Delta u(r,\theta)=\Delta K \frac{1+\nu}{E}\sqrt{\frac{r}{2\pi}}\cos\left(\frac{\theta}{2}\right)(3-4\nu-\cos\theta)
    \label{despu}
\end{equation}
\begin{equation}
   \Delta  v(r,\theta)= \Delta K \frac{1+\nu}{E}\sqrt{\frac{r}{2\pi}}\sin\left(\frac{\theta}{2}\right)(3-4\nu-\cos\theta)
    \label{despv}
\end{equation}

\noindent where $r$ and $\theta$ denote the radial and angular coordinates of a polar coordinate system centred at the crack tip. Plane strain conditions and finite deformations are considered. An initial crack tip blunting radius is defined $r_0=0.5$ $\mu$m, rendering an initial crack tip opening displacement of $b_0=1$ $\mu$m \cite{Dadfarnia2011}. The outer radius is chosen to be 300,000 times larger than $r_0$. As depicted in Fig. \ref{fig:Boundary}, cyclic loading is imposed by scaling in time $t$ the external load by a sinusoidal function with amplitude $\Delta K = K_{max} - K_{min}$ and ratio $R=K_{min}/K_{max}$. A load ratio of $R=0.1$ is used throughout the study and the number of cycles is denoted by $N$. Mimicking the experiments, an initial hydrogen concentration $C_0$ is prescribed uniformly in the entire sample. We will also model the case of an open-system, where we prescribe a constant chemical potential, as described in Section \ref{sec:ConvFreq} below. Mechanical and diffusion properties are those measured in Section \ref{Sec:Material}, with the partial molar volume of hydrogen taken to be $\bar{V}_H=2 \times 10^{-6}$ m$^3$/mol.

\begin{figure}[H]
  \makebox[\textwidth][c]{\includegraphics[width=0.7\textwidth]{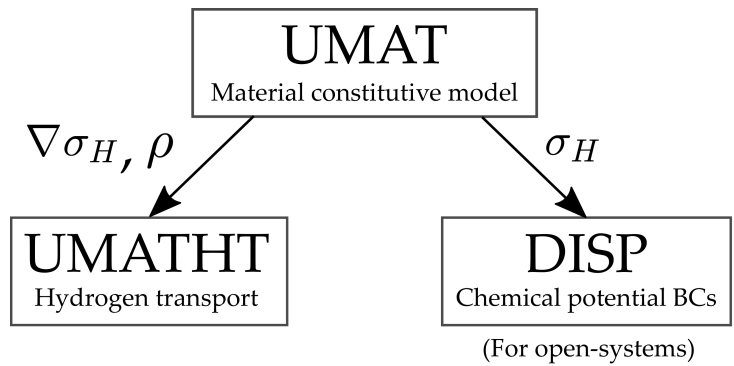}}
  \caption{Flow chart describing the coupling between the user subroutines employed in the numerical implementation.}
  \label{fig:Flow}
\end{figure}

The multi-trap hydrogen transport and micromechanics constitutive models described in Section \ref{Sec:Theory} are implemented in the commercial finite element package Abaqus using, respectively, a UMATHT and a UMAT subroutine. A DISP subroutine is employed to prescribe a constant chemical potential at the crack faces \cite{IJHE2016,Diaz2016b}. The coupling between the different user subroutines is described in Fig. \ref{fig:Flow}. The model is discretised using 5238 quadrilateral quadratic elements with reduced integration. The use of a finer mesh leads to convergence problems for high values of $\Delta K$ due to large element distortions. However, at low $\Delta K$ values, the present mesh appears to give results that are quantitatively similar to those obtained with finer meshes.

\section{Results}
\label{Sec:Results}

The influence of cyclic loading on the lattice and trapped hydrogen distributions is investigated first, Section \ref{Sec:TrapANDLattice}. We then proceed, in Section \ref{sec:ConvFreq}, to shed light into the role of frequency and boundary conditions (closed-system versus open-system). In Section \ref{Sec:BeneficialTraps} we quantify the implications of engineering alloys with a higher density of carbide trapping sites. Finally, the sensitivity of the results to reliable measurements of the binding energy and the role of plastic strain gradients are respectively addressed in Sections \ref{Sec:BindingEnergy} and \ref{Sec:CMSG}.

\subsection{Cyclic behaviour of trap and lattice concentrations}
\label{Sec:TrapANDLattice}

\begin{figure}[H]
  \makebox[\textwidth][c]{\includegraphics[width=1.3\textwidth]{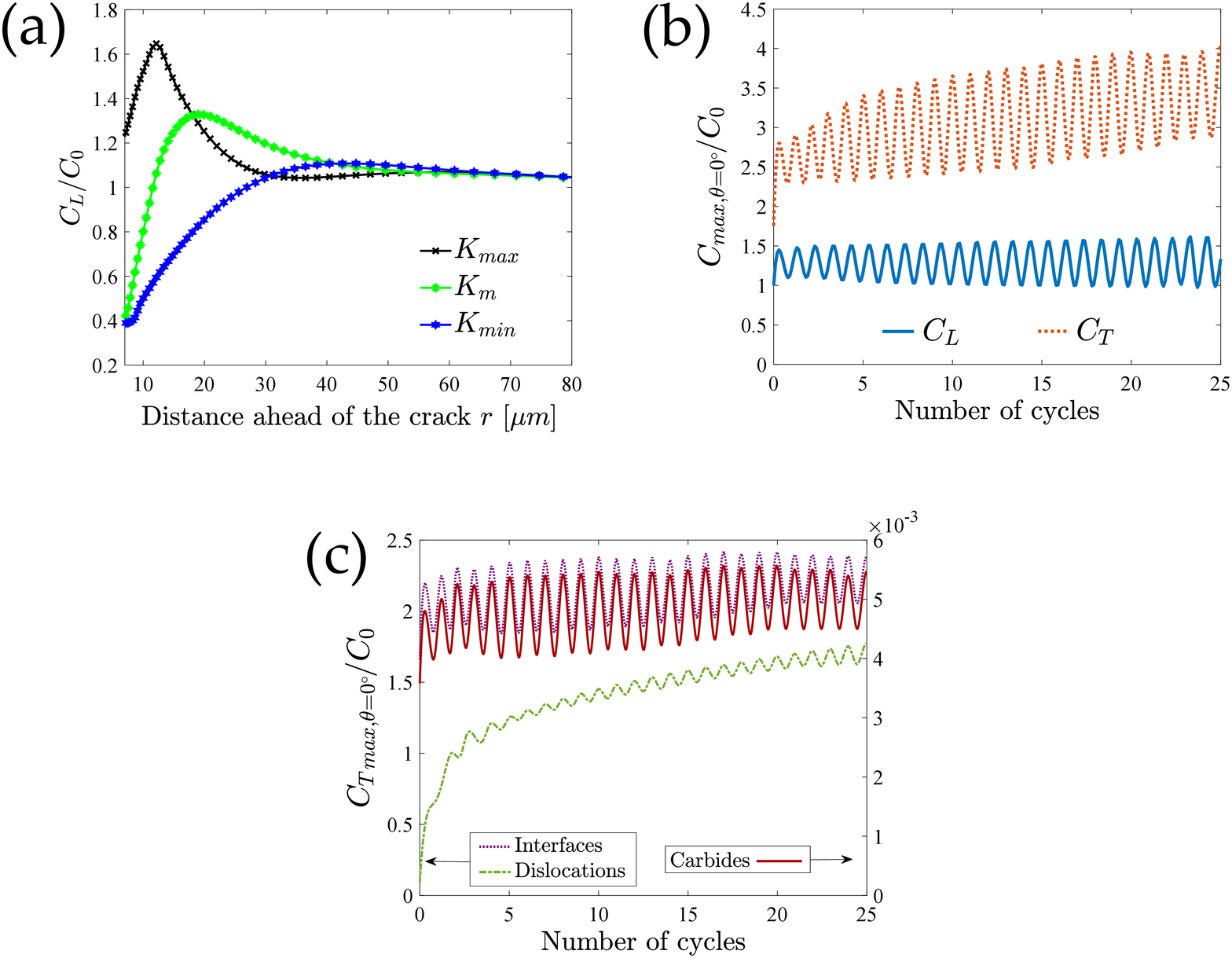}}
  \caption{Cyclic behaviour of the hydrogen concentrations; (a) lattice hydrogen distribution ahead of the crack for the 10th cycle, and time/cycle dependency of the maximum lattice and trapped hydrogen concentration: (b) summed contribution from all trap sites, and (c) individual contributions. Fatigue loading, $f=1$ Hz, $R=0.1$, $\Delta K=35$ MPa$\sqrt{m}$.}
  \label{fig:TrapLattice}
\end{figure}

Consider the 42CrMo4 steel characterised in Section \ref{Sec:Material}. The behaviour of lattice $C_L$ and trapped $C_T$ hydrogen concentrations are shown in Fig. \ref{fig:TrapLattice} for a sample pre-charged with $C_0=1.06$ wt ppm, as in the experiments. A frequency of $f=1$ Hz is considered, as this corresponds to the frequency level at which the same experimental response is observed with and without hydrogen, and $\Delta K=35$ MPa$\sqrt{m}$. First, Fig. \ref{fig:TrapLattice}a shows the lattice hydrogen concentration at three different stages of a representative cycle ($N$=10): the maximum load $K_{max}$, the minimum load $K_{min}$ and the mean load $K_m=K_{max}-|K_{min}|$. In agreement with expectations, the hydrogen concentration follows qualitatively the trend depicted by the applied load, with the three curves merging far away from the crack tip (where $\sigma_H$ is small). The variation of the maximum values of $C_L$ and $C_T$ ahead of the crack ($\theta=0^{\circ}$) are shown in Fig. \ref{fig:TrapLattice}b as a function of time (number of cycles), where $C_T$ includes the contributions from all trap sites. This quantity, denoted ${C_L}_{max,\theta=0^\circ}$ (or ${C_T}_{max,\theta=0^\circ}$), is the maximum magnitude of $C_L$ (or $C_T$) attained for a given instant of time across all material points ahead of the crack tip; i.e., ${C_L}_{max,\theta=0^\circ}=\text{max}(C_L(r,\theta=0^{\circ},t))$. Consistent with Fig. \ref{fig:TrapLattice}a and Oriani's equilibrium, the results in Fig. \ref{fig:TrapLattice}b reveal a cyclic variation of ${C_L}_{max,\theta=0^\circ}$ and ${C_T}_{max,\theta=0^\circ}$. The lattice hydrogen concentration exhibits an almost periodic response while the trapped hydrogen concentration increases with time. The trend depicted by $C_T$ is due to the evolution of the dislocation trap density $N_T^{(d)}$ with plastic deformation; this is shown in Fig. \ref{fig:TrapLattice}c, where the individual contributions of each trap type are plotted. More hydrogen is trapped in martensitic interface trapping sites as the trap density is substantially higher, see Table \ref{tab:energy}.  

We will draw implications for embrittlement by focusing on the maximum lattice concentration ${C_L}_{max}$. Oriani's equilibrium provides a one-to-one relation between the lattice and trapped hydrogen concentrations and accordingly $C_L$ can be used to construct a unique failure locus \cite{Ayas2014}. In this way, we refrain from making any mechanistic choices regarding the damage process, retaining the generality of the analysis. 

\subsection{Influence of frequency and boundary conditions}
\label{sec:ConvFreq}

We proceed to shed light into the influence of the loading frequency and the hydrogen charging conditions. Dimensional analysis shows that the role of the frequency scales with the diffusion coefficient, such that a normalised frequency can be defined as
\begin{equation}\label{eq:f_bar}
    \bar{f} = \frac{f R_p^2}{D_e} \, ,
\end{equation}

\noindent where $R_p$ is the fracture process zone, given by the Irwin approximation as
\begin{equation}\label{eq:Rp}
    R_p = \frac{1}{3 \pi} \left( \frac{K_I}{\sigma_y} \right)^2 \, .
\end{equation}

\noindent For simplicity, we define $\bar{f}$ using $D_L$, as it remains constant throughout the analysis, and use $K_{max}$ in (\ref{eq:Rp}). Accordingly, a normalised time can be given by $\bar{t}=D_L t/R_p^2$. The evolution of the maximum lattice hydrogen concentration is computed for different frequencies in two scenarios: a closed-system, where the sample is pre-charged with $C_0$, and an open-system, where the sample is pre-charged with $C_0$ and continuously exposed to a permanent source of hydrogen. In the latter case, the appropriate boundary condition in the crack faces is to prescribe a constant chemical potential \cite{DiLeo2013}. Using the concentration as a degree of freedom, this equates to the following boundary condition
\begin{equation}
    C_b = C_0 \exp \left( \frac{\bar{V}_H \sigma_H}{RT} \right) \, .
\end{equation}

\begin{figure}[H]
  \makebox[\textwidth][c]{\includegraphics[width=1.\textwidth]{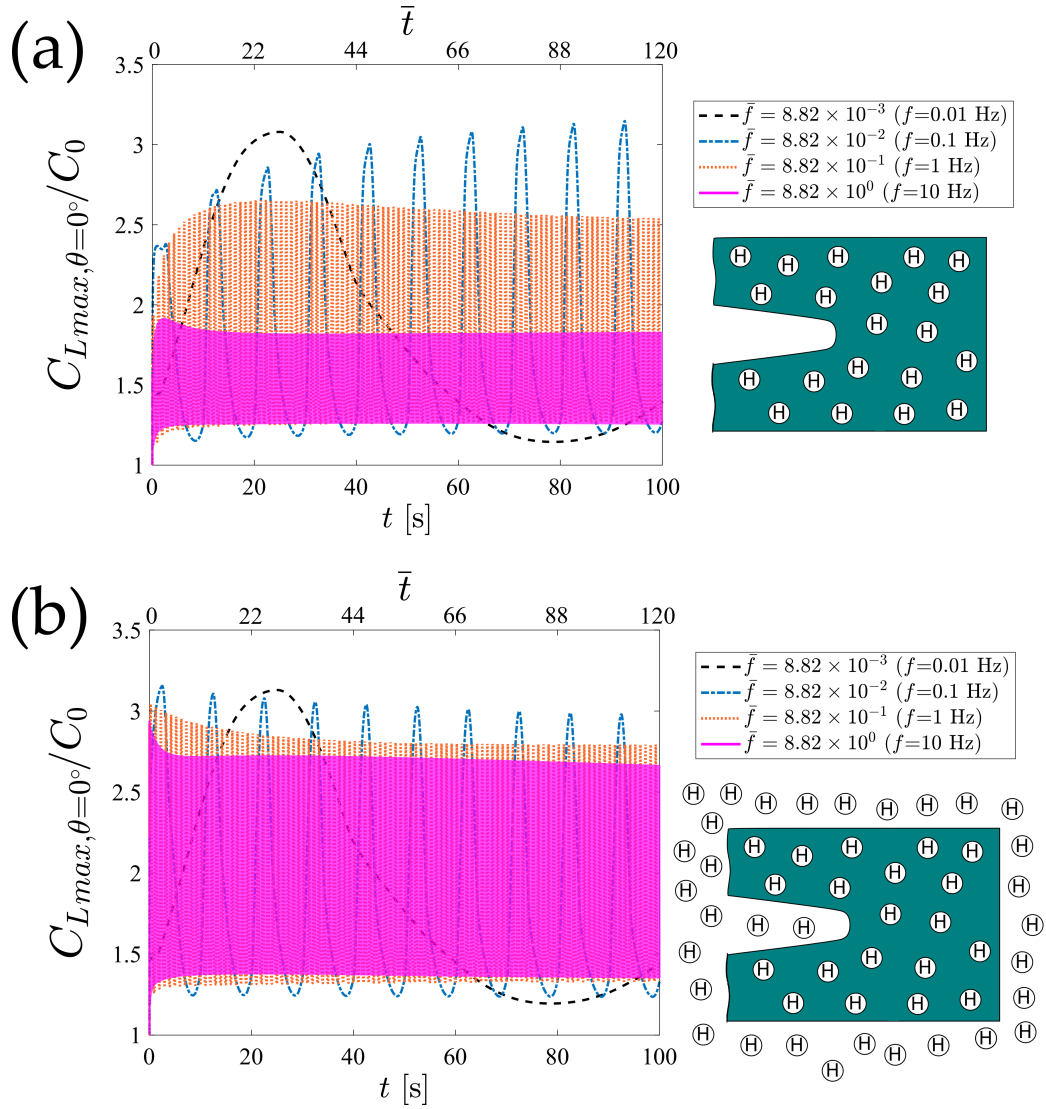}}
  \caption{Influence of loading frequency; variation in time of the maximum value of $C_L$ ahead of the crack for: (a) a closed-system, and (b) an open-system. Fatigue loading, $R=0.1$, $\Delta K=10$ MPa$\sqrt{m}$.}
  \label{fig:Frequency}
\end{figure}

The results are shown in Fig. \ref{fig:Frequency}. In both closed- and open-systems the hydrogen concentration follows the cyclic behaviour of the applied load. The maximum value attained by $C_L$ is practically constant after a number of cycles. Some humps are observed for the highest frequencies as a larger number of cycles is considered. However, based on the crack growth rates of Fig. \ref{fig:Propagation}, crack extension is likely to occur at a lower number of cycles, re-distributing crack tip fields. More importantly, the magnitude of the maximum $C_L$ attained depends on the $f/D_e$ ratio; if we load at high frequencies or use materials with low (effective) diffusion coefficients, ${C_L}_{max,\theta=0^\circ}$ will be lower. This is quantified as a function of time in Figs. \ref{fig:Frequency}a and \ref{fig:Frequency}b for closed and open-systems, respectively. The results show that the magnitude of ${C_L}_{max,\theta=0^\circ}$ varies cyclically and shows a high sensitivity to $\bar{f}$; critical values of the hydrogen concentration may not be attained if the loading frequency is sufficiently high or $D_e$ is sufficiently low. It is important to emphasize that this behaviour is observed for both closed- and open-systems, Figs. \ref{fig:Frequency}a and \ref{fig:Frequency}b respectively, albeit to a lesser degree in the latter. Calculations conducted for open-systems with the same $C_b$ but a smaller (or zero) initial hydrogen concentration exhibit the same qualitative trends, although the effect is smaller. Moreover, the observed sensitivity of the maximum value of $C_L$ to the loading frequency is also the expected qualitative behaviour at large time scales (number of cycles). If $D_e$ is small relative to the time required to complete one loading cycle, the steady state behaviour of $C_L$ will not be governed by the maximum value of $\sigma_H$ but by the mean. In other words, the capacity of the hydrogen distribution to reach its upper limit, given by the steady state solution for the $\sigma_H$ distribution associated with $K_{max}$, is governed by the ratio between the loading frequency $f$ and the effective diffusion coefficient $D_e$. The implications are profound, if alloys can be engineered to reduce the effective diffusion coefficient, resistance to hydrogen assisted fatigue can be gained over a larger range of loading frequencies and in all applications. This will be quantified in the following Section.

\subsection{Hydrogen-trap interaction: can it be used to mitigate fatigue?}
\label{Sec:BeneficialTraps}

\begin{figure}[H]
  \makebox[\textwidth][c]{\includegraphics[width=1.1\textwidth]{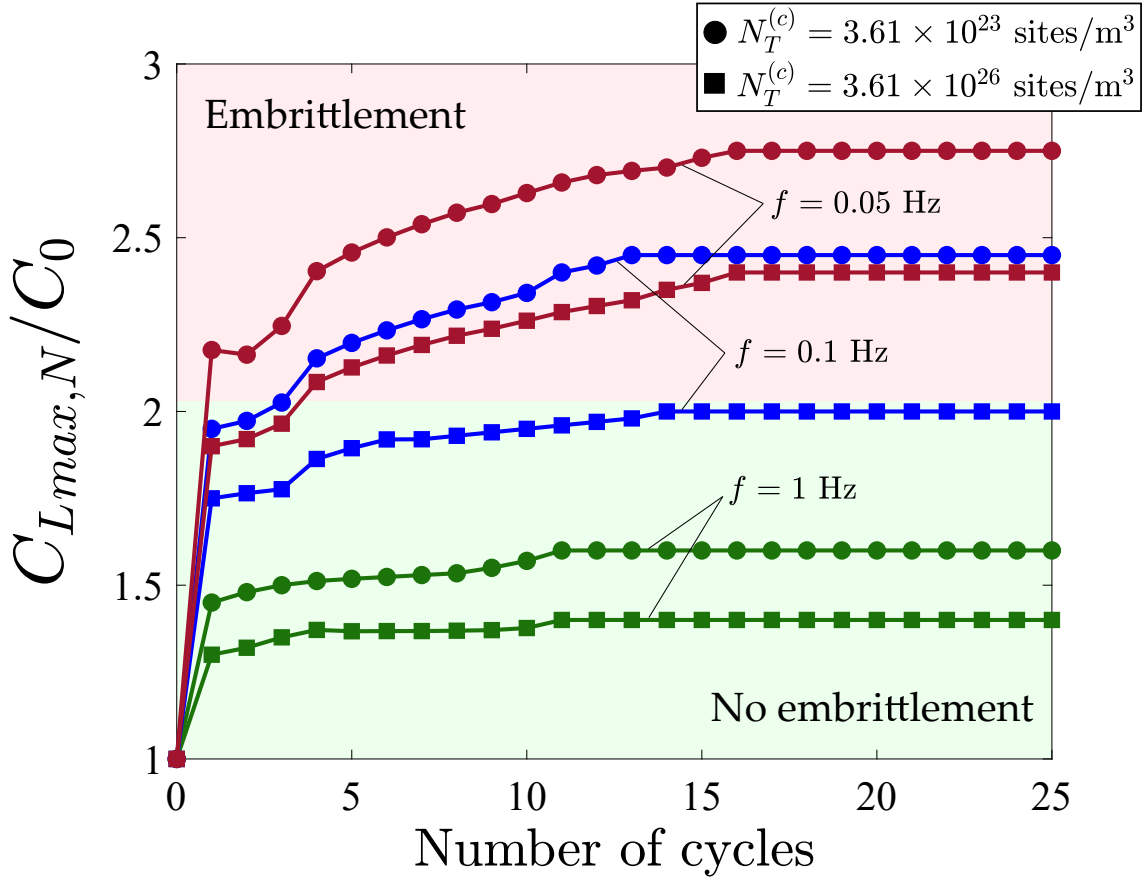}}
  \caption{Influence of increasing the carbide trap density $N_T^{(c)}$. Maximum $C_L$ in each cycle as a function of time (number of cycles) for three frequencies and two carbide trap densities. Fatigue loading, $R=0.1$, $\Delta K=35$ MPa$\sqrt{m}$. The region of embrittlement is inferred from the experimental results on 42CrMo4 steel.}
  \label{fig:BeneficialTraps}
\end{figure}

Let us gain quantitative insight by correlating with the experiments described in Section \ref{Sec:Material}. Fig. \ref{fig:Frequency} reveals that the maximum value attained by $C_L$ along the extended crack plane rapidly reaches a plateau in time, with the magnitude of this plateau value being highly sensitive to the frequency. As discussed in Section \ref{Sec:TrapANDLattice}, we will assume that a threshold value for $C_L$ exists that determines the onset of embrittlement. The maximum value of $C_L$ attained in each cycle is plotted in Fig. \ref{fig:BeneficialTraps} for the three loading frequencies considered in the experiments. This quantity, estimated once per cycle, is denoted as ${C_L}_{max,N}$ to differentiate from ${C_L}_{max,\theta=0^\circ}$ (the maximum value of $C_L$ at each time instant, which varies cyclically). Recall that, as shown in Fig. \ref{fig:Propagation}, a loading frequency of 1 Hz or higher does not lead to any embrittlement while an increase in fatigue crack growth rates can be observed for frequencies of 0.1 Hz or lower. Thus, the critical value of $C_L$ at which embrittlement is observed must be within the plateau values of ${C_L}_{max,N}$ predicted for $f=0.1$ and $f=1$ Hz. Results for the \emph{standard} carbide trap density, $N_T^{(c)}=3.61 \times 10^{23}$ sites/m$^3$, reveal plateau values of 2.5$C_0$ and 1.6$C_0$ for, respectively $f=0.1$ and $f=1$ Hz. Taking the average, we stipulate that the critical $C_L$ for embrittlement equals 2.05$C_0$, as depicted in Fig. \ref{fig:BeneficialTraps}. 

Fig. \ref{fig:BeneficialTraps} also includes results obtained assuming that the density of carbides can be engineered. Namely, the density of carbide trapping sites is increased to $N_T^{(c)}=3.61 \times 10^{26}$ sites/m$^3$ \cite{Ramjaun2018}. 
Increasing the trap density decreases the diffusion coefficient, see (\ref{eq:De}), and accordingly, the diffusion of lattice hydrogen within each cycle is reduced, leading to a lower value of ${C_L}_{max,N}$. As shown in Fig. \ref{fig:BeneficialTraps}, the maximum value of hydrogen concentration attained with $f=0.05$ Hz in an alloy with additional traps is similar to that obtained in a \emph{standard} alloy for a frequency of $f=0.1$ Hz. Moreover, the simulations for the trap-enhanced material show no embrittlement within the $f=0.1-1$ Hz regime, effectively extending the regime of safe frequencies at which hydrogen has no effect by an order of magnitude. 

\subsection{Influence of the binding energy}
\label{Sec:BindingEnergy}

\begin{figure}[H]
  \makebox[\textwidth][c]{\includegraphics[width=1.1\textwidth]{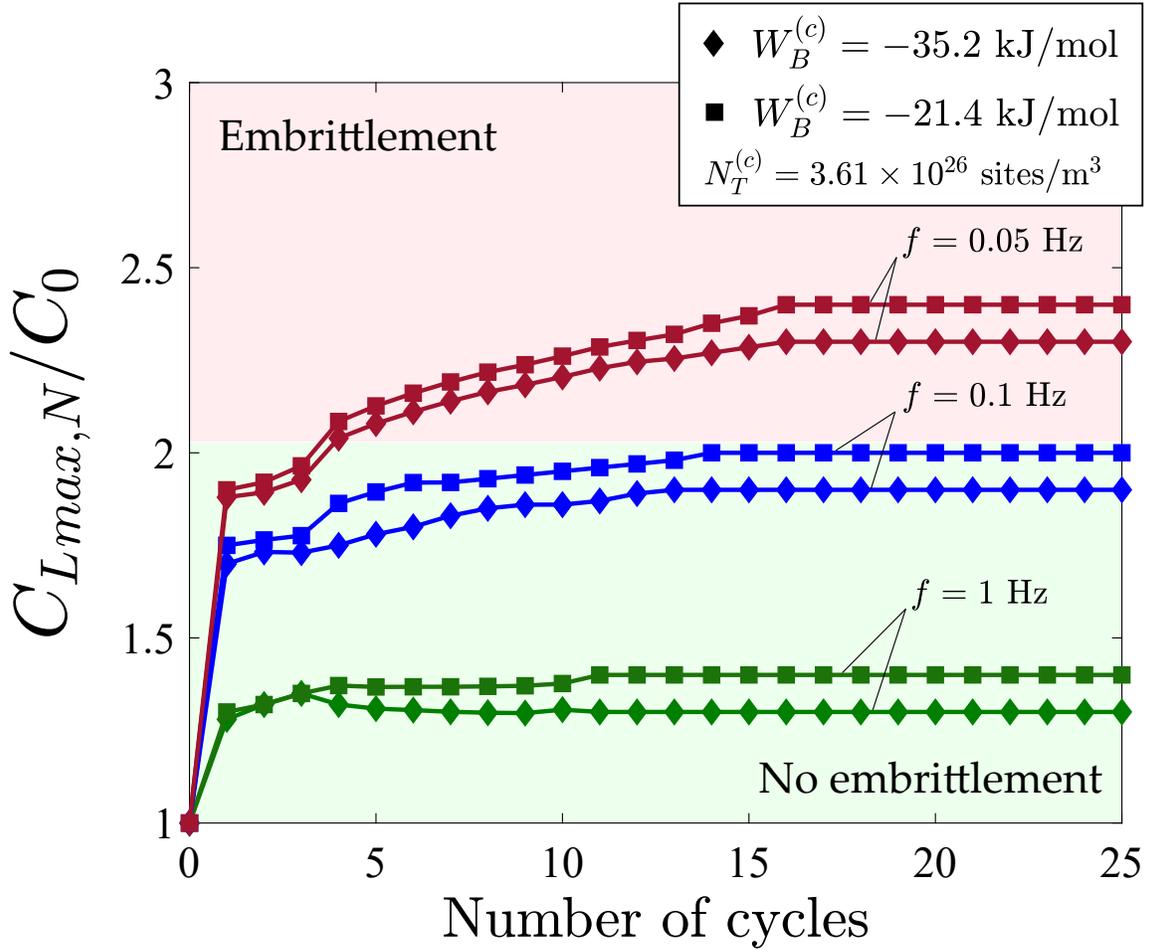}}
  \caption{Influence of the binding energy. Maximum $C_L$ in each cycle as a function of time (number of cycles) for three frequencies and two carbide binding energies in a high carbide trap density material. Fatigue loading, $R=0.1$, $\Delta K=35$ MPa$\sqrt{m}$. The region of embrittlement is inferred from the experimental results on 42CrMo4 steel.}
  \label{fig:InfluenceBindingEnergy}
\end{figure}

Traps are characterised by the binding energy and the trap density. In the following, we examine the implications of the binding energy $W_B$ in the above conclusions. The inverse problem of interpreting TDS data to determine $W_B$ for each trap type is ill-posed, bringing uncertainties into the analysis. Thus, we consider that carbides are the strongest trap in the model, exchanging the value of its binding energy with that of dislocations ($W_B=-35.2$ kJ/mol, see Table \ref{tab:energy}). The results are shown in Fig. \ref{fig:InfluenceBindingEnergy} assuming a high carbide trap density material, $N_T^{(c)}=3.61 \times 10^{26}$ sites/m$^3$. Predictions are compared to those obtained with the previous binding energy estimate $W_B^{(c)}=-21.4$ kJ/mol; the higher $|W_B|$, the lower the maximum value of $C_L$ attained. The results do not exhibit the high sensitivity shown by the trap density but the variation in $W_B$ is also low (13.8 kJ/mol). Given that $21.4$ kJ/mol is on the lower side of the $|W_B|$ values reported for carbides \cite{Song2015}, it is expected that the conclusions drawn in the previous section will hold to a greater degree. Moreover, results suggest that the gains derived from an increase in trap density could be significantly enhanced if the density of deep traps $|W_B| > 50$ kJ/mol is increased.

\subsection{Influence of plastic strain gradients}
\label{Sec:CMSG}

\begin{figure}[H]
  \makebox[\textwidth][c]{\includegraphics[width=1.35\textwidth]{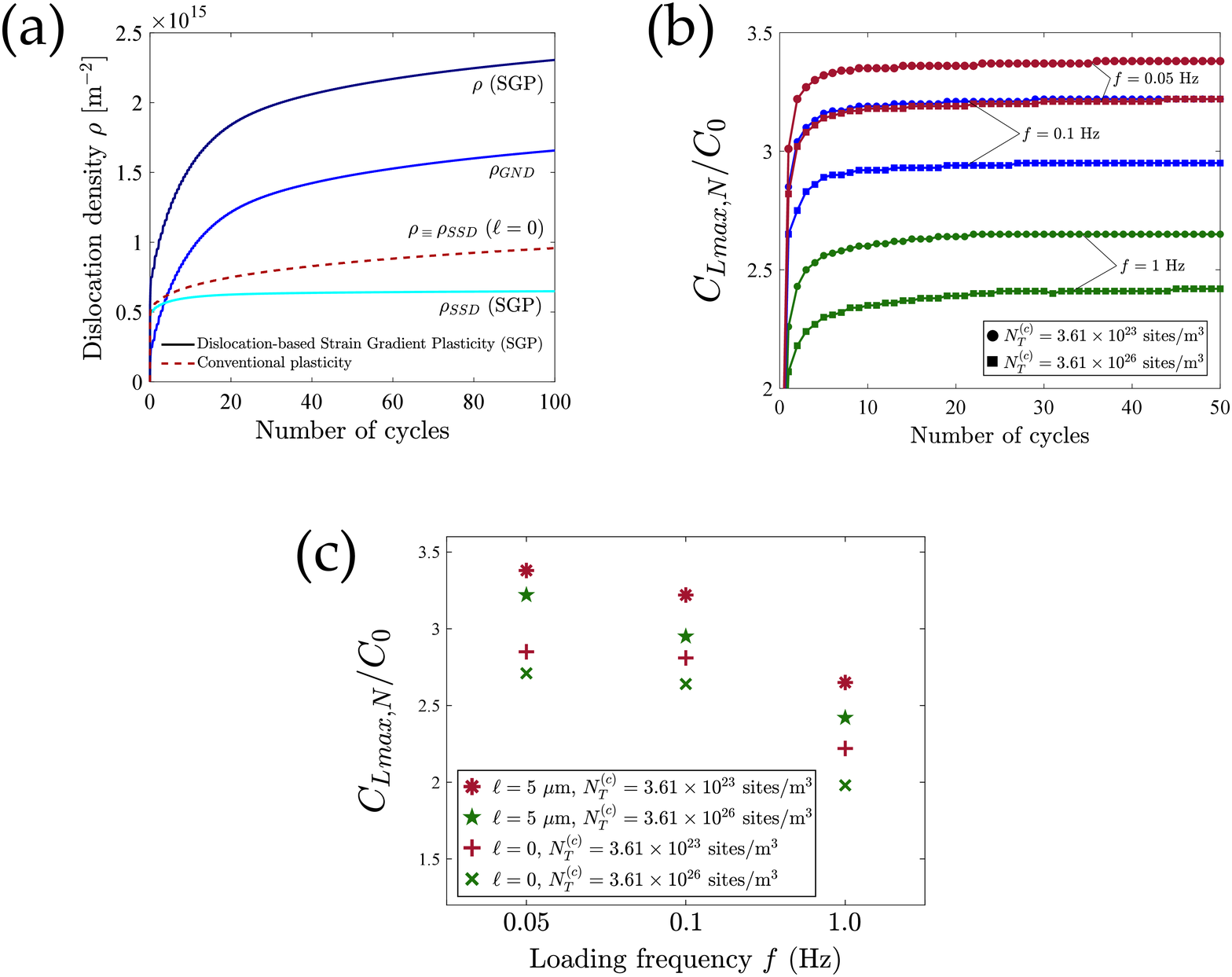}}
  \caption{Influence of plastic strain gradients; (a) dislocation density evolution with time (number of cycles) for conventional and strain gradient plasticity ($f=1$ Hz), (b) maximum $C_L$ in each cycle as a function of time (number of cycles) for three frequencies and two carbide trap densities, (c) maximum value attained by $C_L$ in the analysis for conventional and strain gradient plasticity, two carbide trap densities and three frequencies. All quantities have been sampled at $x_{crit}=2$ $\mu$m from the crack tip. Fatigue loading, $R=0.1$, $\Delta K=10$ MPa$\sqrt{m}$.}
  \label{fig:SGP}
\end{figure}

To ease the interpretation of the results, conventional $J_2$ plasticity has been used for the computations reported so far. However, it has been shown that plastic strain gradients are very high near the crack tip and elevate dislocation density and local strength \cite{Wei1997,Komaragiri2008,IJSS2015}. This local dislocation hardening associated with GNDs can be accounted for using strain gradient plasticity theories \cite{Fleck1994,Gao1999,Anand2005,JMPS2019}. Considering the dislocation-based strain gradient plasticity model formulated in Section \ref{Sec:Theory}, two implications can be foreseen. First, common to other gradient plasticity models, crack tip stresses will be significantly higher than those predicted by conventional plasticity, which in turn increases the lattice hydrogen content close to the crack tip \cite{IJHE2016}. Secondly, total dislocation density $\rho$ predictions (and the associated trap density $N_T^{(d)}$) can differ significantly. It is expected that the statistically stored dislocation (SSD) density $\rho_{SSD}$, predicted via (\ref{eq:rhoS}), will be lower than in conventional plasticity as local hardening reduces $\varepsilon^p$, but there will be an additional contribution to the total dislocation density as GNDs are accounted for ($\rho_{GND}$). The influence of these features on the cyclic behaviour of lattice and trapped hydrogen concentration are investigated here for the first time. Note that, unlike conventional plasticity, strain gradient plasticity predictions do not predict a peak $\sigma_H$ (and $C_L$) at a finite distance ahead of the crack tip \cite{AM2016,EJMAS2019}; accordingly, we do not compute the maximum value of the variables under consideration along the crack ligament but sample them at a critical distance for embrittlement. As shown by Gangloff \cite{Gangloff2003a,Gangloff1990} using $D_e$ and stage II $da/dt$ data, this critical distance is of a few microns in many alloys, which further motivates the use of strain gradient plasticity models; $x_{crit}=2$ $\mu$m is assumed. Also, the material gradient length scale is assumed to be equal to $\ell=5$ $\mu$m, an intermediate value within the range of length scales reported in the literature from micro-scale experiments \cite{IJES2020}. 
Fig. \ref{fig:SGP}a shows the evolution of the dislocation density with the number of loading cycles, where the maximum value of $\rho_i$ at $x_{crit}=2$ $\mu$m is plotted. As expected, local strain gradient hardening reduces $\rho_{SSD}$ and the density of statistically stored dislocations is higher in the conventional plasticity case ($\ell=0$). However, the total dislocation density $\rho$ is substantially higher in the case of the strain gradient plasticity model as the GND density is notably larger than the total dislocation density predicted with conventional plasticity. This, in turn, translates into a significantly higher trap density for dislocation sites $N_T^{(d)}$ as the number of cycles increases. We proceed to examine if the conclusions drawn in previous sections are still applicable in view of this notably different crack tip behaviour. As with the case of conventional plasticity, the maximum value of $C_L$ per cycle remains practically constant after a certain number of cycles. Fig. \ref{fig:SGP}b shows the variation in time of the maximum value of $C_L$ per cycle at $x_{crit}$ for strain gradient plasticity, three selected values of the loading frequency and two carbide trap densities. The qualitative trends are the same as those obtained so far with conventional plasticity; increasing the density of carbide trapping sites reduces the maximum lattice hydrogen concentration attained. A quantitative comparison with the results from conventional plasticity is shown in Fig. \ref{fig:SGP}c. Strain gradient plasticity predicts a higher value of the maximum hydrogen concentration in all cases due to the higher crack tip stresses, this would translate into a higher experimentally-calibrated critical $C_L$ for embrittlement. The drop in $C_{Lmax}$ with increasing $N_T^{(c)}$ is quantitatively similar to that predicted with conventional plasticity. Therefore, the use of more accurate micromechanics-based descriptions of crack tip fields does not change the conclusions drawn before with conventional plasticity.

\section{Conclusions}
\label{Sec:ConcludingRemarks}

We have presented a micromechanics-based multi-trap model for stress-assisted hydrogen diffusion. The model is used to investigate the competing role of the loading frequency $f$ and the effective diffusion coefficient $D_e$ on hydrogen assisted fatigue in the presence of multiple microstructural traps. Experiments on 42CrMo4 steel are used to gain quantitative insight by inferring a critical hydrogen concentration for embrittlement based on the frequency range where hydrogen has no effect on crack growth rates. The main findings are:\\

\noindent (i) The trap and lattice hydrogen concentration vary cyclically following the variation of the mechanical load. The maximum concentration value attained in each cycle by the hydrogen trapped at dislocations rises with time as the associated trap density $N_T^{(d)}$ increases with plastic deformation. Contrarily, the maximum values of the lattice hydrogen concentration $C_L$ and the hydrogen trapped at other traps such as carbides or interfaces remain practically constant after a few cycles.\\

\noindent (ii) The maximum hydrogen concentration attained ahead of the crack is highly sensitive to the $D_e/f$ ratio. A lower peak in the $C_L$ distribution is observed if the effective diffusion coefficient is small relative to the time required to complete a loading cycle. This behaviour is observed in both closed-systems (one-off hydrogen entry) and open-systems (permanent source of hydrogen). \\

\noindent (iii) Increasing the density of ``beneficial'' traps not involved in the fracture process is a viable strategy for mitigating hydrogen assisted fatigue. An increase in the density of carbide trapping sites $N_T^{(c)}$ reduces the maximum hydrogen concentration values attained for a given frequency and extends the regime of \emph{safe} frequencies where embrittlement is not predicted.\\

\noindent (iv) The maximum concentration values predicted for a given frequency can be further reduced by increasing the density of the trapping sites with stronger binding energy $|W_B|$. Quantitatively, the effect is lower than varying the trap density, as the range of binding energies is more limited. \\

\noindent (v) The use of strain gradient plasticity to better resolve crack tip fields shows a higher dislocation trap density and lattice hydrogen concentration relative to conventional plasticity predictions. However, no significant qualitative or quantitative differences are observed regarding the role of an increased carbide trap density in mitigating hydrogen assisted fatigue.\\

Given that most engineering components are subjected to cyclic loads, these insights could have important implications in the design of hydrogen-resistant alloys. 

\section{Acknowledgements}
\label{Sec:Acknowledgeoffunding}

The authors would like to acknowledge helpful discussions with F.J. Belzunce and A. Zafra (University of Oviedo) in regards to the experiments and the trap density measurements. The authors acknowledge funding from the Regional Government of Asturias (grant FC-GRUPIN-IDI/2018/000134) and the IUTA (grant SV-19-GIJON-1-19). E. Mart\'{\i}nez-Pa\~neda also acknowledges financial support from EPSRC funding under grant No. EP/R010161/1 and from the UKCRIC Coordination Node, EPSRC grant number EP/R017727/1, which funds UKCRIC's ongoing coordination.






\bibliographystyle{elsarticle-num}
\bibliography{library}

\end{document}